\documentclass[twocolumn]{aastex631}

\newcommand{\Teff}{T$_{\rm{eff}}$}
\newcommand{\ms}{m s$^{-1}$}

\usepackage{graphicx}

\shorttitle{Hic Sunt Dracones: Uncovering Dynamical Perturbers Within
  the Habitable Zone}
\shortauthors{Stephen R. Kane \& Jennifer A. Burt}

\begin{document}

\title{Hic Sunt Dracones: Uncovering Dynamical Perturbers Within the
  Habitable Zone}

\author[0000-0002-7084-0529]{Stephen R. Kane}
\affiliation{Department of Earth and Planetary Sciences, University of
  California, Riverside, CA 92521, USA}
\email{skane@ucr.edu}

\author[0000-0002-0040-6815]{Jennifer A. Burt}
\affil{Jet Propulsion Laboratory, California Institute of Technology,
  4800 Oak Grove Drive, Pasadena, CA 91109, USA}


\begin{abstract}

The continuing exploration of neighboring planetary systems is
providing deeper insights into the relative prevalence of various
system architectures, particularly with respect to the solar
system. However, a full assessment of the dynamical feasibility of
possible terrestrial planets within the Habitable Zones (HZ) of nearby
stars requires detailed knowledge of the masses and orbital solutions
of any known planets within these systems. Moreover, the presence of
as yet undetected planets in or near the HZ will be crucial for
providing a robust target list for future direct imaging surveys. In
this work, we quantify the distribution of uncertainties on planetary
masses and semi-major axes for 1062 confirmed planets, finding median
uncertainties of 11.1\% and 2.2\%, respectively. We show the
dependence of these uncertainties on stellar mass and orbital period,
and discuss the effects of these uncertainties on dynamical analyses
and the locations of mean motion resonance. We also calculate the
expected radial velocity (RV) semi-amplitude for a Neptune-mass planet
in the middle of the HZ for each of the proposed Habitable Worlds
Observatory target stars. We find that for more than half of these
stars, the RV semi-amplitude is less than 1.5 \ms\, rendering them
unlikely to be detected in archival RV data sets and highlighting the
need for further observations to understand the dynamical viability of
the HZ for these systems. We provide specific recommendations
regarding stellar characterization and RV survey strategies that work
toward the detection of presently unseen perturbers within the HZ.

\end{abstract}



\section{Introduction}
\label{intro}

The number of known exoplanetary systems now number in the thousands,
enabling statistical studies of exoplanet demographics as well as
detail characterization of individual systems
\citep{ford2014,winn2015,he2019}. Improvements in radial velocity (RV)
precision combined with long-term survey strategies are gradually
filling out the wider-separation system architectures for the nearest
and brightest stars, revealing giant planets beyond the snow line
\citep{fischer2016,wittenmyer2020b,fulton2021,rosenthal2021} and
providing comparisons to the orbital configuration of the solar system
\citep{horner2020b,kane2021d}. However, there remain significant
limitations on how well existing observations can probe for low-mass
planets in the Habitable Zone (HZ) of these stars, many of which will
serve as targets for future observations that focus on potentially
habitable worlds
\citep{kasting1993a,kane2012a,kopparapu2013a,kopparapu2014,chandler2016,kane2016c,hill2018,hill2023}.
Furthermore, improving the orbital parameter precision of known
exoplanets and their host stars is of particular importance for
enabling follow-up observations and system analyses
\citep{ford2005a,kane2009c}. For example, reduced uncertainties in
planetary masses are crucial for robust interpretation of atmospheric
spectra \citep{batalha2017a,batalha2019c,changeat2020a,dimaio2023} and
interior models \citep{unterborn2016,muller2020b}, whilst stellar
uncertainties can impact the correct determination of HZ boundary
locations \citep{kane2014a,kane2018a} and of a rocky planet's core
mass fraction \citep{schulze2021}.  The continued observations of
nearby stars to concisely catalog their planetary inventories is thus
a necessary endeavor for enabling subsequent characterization studies
of these systems.

A coming evolution of exoplanet detection and exploration is that of
directly imaging facilities that will prioritize the detection and
characterization of the aforementioned HZ terrestrial planets
\citep{brown2015a,barclay2017,kane2018c,kopparapu2018,stark2020,li2021a}. Of
particular note are the Habitable Worlds Observatory (HWO) and the
Large Interferometer For Exoplanets (LIFE) mission, currently under
various development phases for consideration as a means to realize
this HZ direct imaging capability
\citep{quanz2022a,vaughan2023,stark2024a}. A major task for the
development of these missions is the construction and refinement of
potential target star lists. An initial list of 164 HWO target stars
has been proposed by \citet{mamajek2024} for discussion and assessment
by the community. Science yield calculations based on these proposed
HWO targets have been calculated by \citet{morgan2023} and
\citet{tuchow2024}, and the stellar properties have been cataloged in
detail by \citet{harada2024b}.

A further consideration is the presence of planets in these systems
that may dynamically impede the presence of stable orbits within the
HZ. A recent study by \citet{kane2024d} found that 11 of the 30 HWO
stars that are known to harbor planets have architectures that
eliminate or drastically diminish the ability of those stars to host
terrestrial HZ planets. In order to further such analyses for these
(and other) stars, continued RV and astrometric measurements must be
carried out to both reduce the uncertainties of the derived planetary
parameters and detect previously unknown planets. This is especially
important for Neptune or sub-Neptune planets within the HZ that,
although lying beneath the sensitivity of most archival RV datasets,
can still act as gravitational perturbers whose presence has dire
consequences for dynamical stability of terrestrial HZ planets.

In this paper, we present the results of an analysis of current
planetary parameter uncertainties, their effect on orbital dynamics,
and the calculated RV signature of potential dynamical perturbers that
may yet be lurking within the HZ of the proposed HWO target stars.  To
borrow from an old Latin phrase, ``Hic sunt dracones", meaning ``Here
be dragons", is an apt phrase when exploring the HZ for objects whose
presence may render the region unsuitable for habitable
environments. In Section~\ref{uncert}, we provide an assessment of the
derived uncertainties for planet mass and semi-major axis, both of
which are key parameters for dynamical simulations, and their
dependence on stellar mass and orbital period measurements. In
Section~\ref{pert}, we describe a proposed HWO target list and compare
the expected RV semi-amplitudes for a Neptune-mass planet in the
center of each star's HZ, along with the challenges in achieving the
necessary RV precision to detect such signatures.
Section~\ref{discussion} provides an additional discussion regarding
the implications of our results and recommendations for a continued
dynamical assessment of stars that are considered as future direct
imaging targets. A summary of our work and concluding remarks are
delivered in Section~\ref{conclusions}.


\section{Derived Planetary System Uncertainties}
\label{uncert}

To investigate the uncertainties in planetary system properties, we
extracted data from the Confirmed Planets
Table\footnote{doi:10.5281/zenodo.11020} of the NASA Exoplanet Archive
\citep{akeson2013} as accessed on 2024 August 22 \citep{nea}. We
selected the default parameter set for each planet included in the
table, and selected only those planets with valid (non-null) values
and uncertainties for the semi-major axis ($a$), planet mass ($M_p$ or
$M_p \sin i$), orbital eccentricity ($e$), orbital period ($P$), and
the mass of the host star ($M_\star$). This provided 1062 planets that
passed our criteria for inclusion in our sample. We note that a small
number of the planets in this list have masses derived from transit
timing variations ($< 4$\%) rather than radial velocity data but we
restrict the overall considerations of this paper to RV-focused
efforts as the transit probability for planets approaching the HZ of
sun-like stars falls off quickly \citep{kane2008b}.


\subsection{Correlations of Uncertainties}
\label{corr}

\begin{figure*}[htb!]
\begin{center}
\includegraphics[width=.98\textwidth]{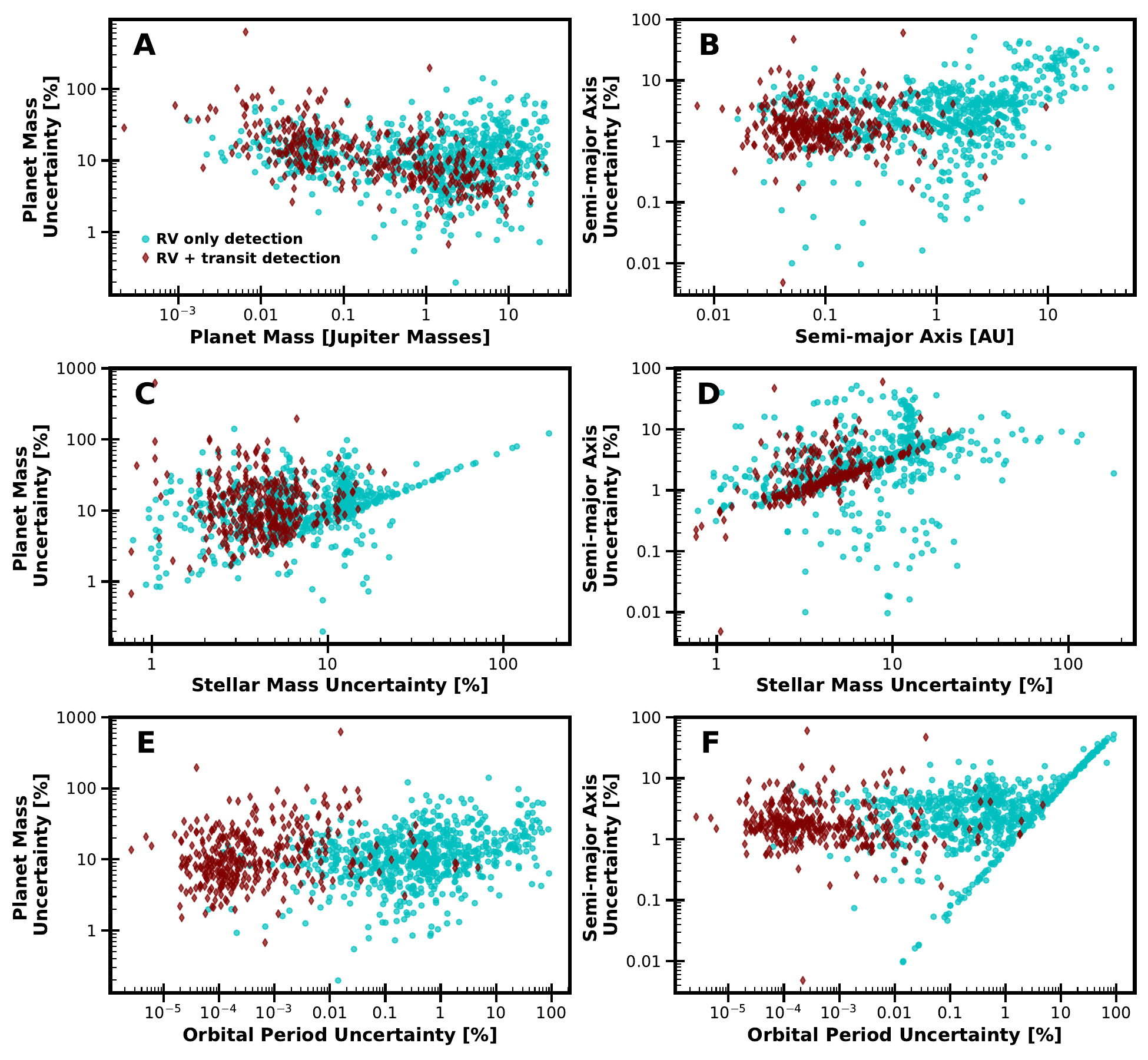}
\caption{Percentage uncertainties for the planetary masses (left
  column) and semi-major axes (right column) of our confirmed planet
  list as a function of their respective parameters (top row; panels A
  and B), stellar mass uncertainties (middle row; panels C and D), and
  orbital period uncertainties (bottom row; panels E and F).  Planets
  detected only in RV surveys are denoted as cyan circles, while those
  detected in both RV and transit surveys are denoted as maroon
  diamonds. The power law relations between stellar and planetary
  masses and the semi-major axis, orbital period, and stellar mass
  (see Equations \ref{eqn:mp} and \ref{eqn:a}) are evident as diagonal
  crowdings of points that set the lower limit in the planet mass and
  semi-major axis uncertainties. For stars with mass uncertainties
  above $\sim$20\%, the stellar mass uncertainty dominates the planet
  mass determination, rather than the precision of the RV
  semi-amplitude measurement. Orbital period uncertainties, however,
  play only a minor role in the planet mass uncertainties. When
  calculating the semi-major axis, orbital period uncertainties beyond
  $\sim$10\% become the dominant source of error but the stellar mass
  uncertainty does not produce a distinct boundary identifying when it
  begins dominating the semi-major axis determination.}
\label{fig:corr1}
\end{center}
\end{figure*}

From our extracted sample of exoplanets, we focused on the two main
planetary properties used as input into dynamical models: the planet
mass and semi-major axis. These properties are generally derived from
other system observables, and their associated uncertainties are
propagated accordingly. In the case of RV observations, the measured
semi-amplitude of the RV variations, $K$, is given by
\begin{equation}
  K = \left( \frac{2 \pi G}{P} \right)^{1/3} \frac{M_p \sin
    i}{(M_p+M_\star)^{2/3}} \frac{1}{\sqrt{1-e^2}}
    \label{eqn:k}
\end{equation}
where the stellar mass, $M_\star$, may be determined via relationships
derived from binary stars \citep{torres2010a} or through a combination
of spectroscopy \citep{yee2017}, asteroseisomology \citep{huber2017d},
and spectral energy distribution analyses
\citep{stassun2018a}. Assuming $M_p \ll M_\star$, the planet mass is
then calculated from the following relation
\begin{equation}
  M_p = \left( \frac{P}{2 \pi G} \right)^{1/3} \frac{M_\star^{2/3} K
    \sqrt{1-e^2}}{\sin i}
  \label{eqn:mp}
\end{equation}
where the orbital period, $P$, and RV semi-amplitude, $K$, are the
primary observables. Similarly, Kepler's third law
\begin{equation}
  P^2 = \frac{4 \pi^2}{G (M_p+M_\star)} a^3
\end{equation}
may be rearranged, assuming $M_p \ll M_\star$, as follows
\begin{equation}
  a = \left( \frac{G M_\star}{4 \pi^2} P^2 \right)^{1/3}
  \label{eqn:a}
\end{equation}
to provide an expression for the semi-major axis, $a$.

In Figure~\ref{fig:corr1}, we provide plots of the percentage
uncertainties for the planetary mass (left column) and semi-major axis
(right column) for all of the planets in our sample. Planets detected
only in RV surveys are denoted as cyan circles, while those detected
in both RV and transit surveys are denoted as maroon diamonds. We
adopt an uncertainty definition of the mean of the absolute values of
the upper and lower uncertainties on a parameter divided by the value
of the parameter itself and then multiplied by 100. The top row plots
these uncertainties as a function of the parameters themselves.  To
provide insight into the dependency of these uncertainties on the
other variables in Equation~\ref{eqn:mp} and Equation~\ref{eqn:a}, we
also plot the uncertainties as a function of the stellar mass
uncertainties (middle row) and orbital period uncertainties (bottom
row).

The planet mass distribution shown in panel A exhibits a well-known
gap within the range 0.1--0.3~$M_J$, first predicted by
\citet{ida2004a} and still a point of discussion as to its
observational significance
\citep{mazeh2016,bennett2021,emsenhuber2021b,hill2023}. The planet
mass uncertainties have a remarkably wide range of 0.2--622\%, but it
should be noted that the planet mass uncertainty outlier at 622\% is
the circumbinary planet Kepler-47 b where the mass uncertainties are
dominated by a relatively unconstrained upper mass limit
\citep{orosz2019}. The planet mass uncertainty distribution has mean
and median values of 16.0\% and 11.1\%, respectively. Panel C shows
that, for stellar mass uncertainties beyond $\sim$20\%, the stellar
mass uncertainty dominates the planet mass determination, below which
it is the precision of the RV semi-amplitude measurement that
dominates. The orbital period uncertainties shown in panel E
demonstrate that the precision of the orbital period measurements play
only a minor role in the planet mass uncertainties.

\begin{figure*}[htb!]
\begin{center}
\includegraphics[width=.98\textwidth]{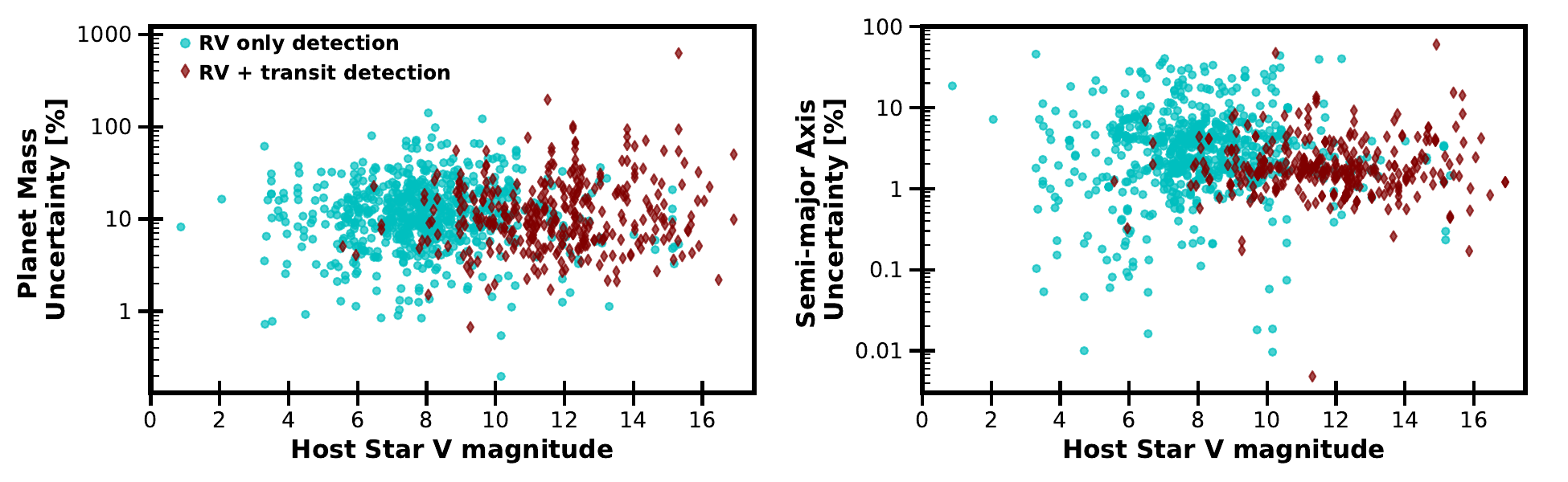}
\caption{Percentage uncertainties for the planetary masses (left
  column) and semi-major axes (right column) of our confirmed planet
  list as a function of host star brightness. Planets detected only in
  RV surveys are denoted as cyan circles, while those detected in both
  RV and transit surveys are denoted as maroon diamonds. The impact of
  transit surveys, which can more efficiently detect planets around
  fainter stars, is notable but neither planetary uncertainty shows a
  significant trend with host star magnitude.}
\label{fig:corr2}
\end{center}
\end{figure*}

The semi-major axes and associated uncertainties are shown in panel B
of Figure~\ref{fig:corr1}. As for the planet mass uncertainties, the
semi-major axis uncertainties span several orders of magnitude,
falling in the range 0.005--60\%, with mean and median values of 4.2\%
and 2.2\%, respectively. For the planets detected in RV surveys only
(cyan circles), the data in panels B, E, and F, show a clear
capability of the RV surveys to detect longer period planets relative
to the transit detections as expected given the decrease in transit
probability with increasing orbital period \citep{kane2008b}. The
semi-major axis uncertainties show a distinct rise at separations
higher than $\sim$5~AU, where the uncertainties become dominated by
the those uncertainties associated with the orbital period rather than
stellar mass. This correlation is validated by the data distributions
shown in panels D and F, where indeed it can be seen that the
uncertainties on semi-major axis correlate broadly with those of
stellar mass, but correlates strongly with those of orbital period in
the high uncertainty regime. The panels of Figure~\ref{fig:corr1} thus
provide a broad overview of the relative contributions toward the
planet mass and semi-major axis uncertainty budgets, and highlight the
importance of improving characterization of host stars and their mass
uncertainties. We further provide the range of percentage uncertainty
values for each parameter discussed here, along with the mean, median,
standard deviation, median absolute deviation (MAD), and interquartile
range (IQR) values, in Table~\ref{tab:stats}. The values shown in the
the table are for all planet masses, and also for the planet mass
groups on either side of 0.1~$M_J$ to emphasize the effects of planet
mass on the measurement precision. Note that the high uncertainty for
semi-major axis and orbital period for the high mass group ($>
0.1$~$M_J$) is the result of the observational bias for high mass
planets at large separations.

\begin{deluxetable*}{l||cccccc}
\tablecolumns{10}
\tablewidth{0pc}
\tablecaption{\label{tab:stats} Parameter percentage uncertainty statistics.}
\startdata 
 & & & & & \\
& \multicolumn{6}{c}{All planet masses} \\ 
Parameter & Range & Mean & Median & Std. Dev. & MAD & IQR\\
\hline
\% Unc. in $M_{p}$&0.2--622.46&16.0&11.1&24.27&5.29&11.8\\
\% Unc. in $M_{\star}$&0.76--181.57&8.4&5.7&10.22&2.93&8.1\\
\% Unc. in $P^\dagger$&0.0--90.16&2.8&0.1&9.35&0.1&0.8\\
\% Unc. in $a$&0.005--60.0&4.2&2.2&6.16&1.36&3.0\\
\hline\hline
& \multicolumn{6}{c}{$M_p \leq 0.1$~$M_J$} \\ 
Parameter & Range & Mean & Median & Std. Dev. & MAD & IQR\\
\hline
\% Unc. in $M_{p}$&1.9-622.46&21.9&15.0&38.4&5.8&12.7\\
\% Unc. in $M_{\star}$&0.76-23.33&5.5&4.5&3.63&1.77&3.6\\
\% Unc. in $P^\dagger$&0.0-6.62&0.2&0.009&0.63&0.01&0.1\\
\% Unc. in $a$&0.005-47.12&2.7&1.7&3.51&0.84&2.4\\
\hline\hline
& \multicolumn{6}{c}{$M_p > 0.1$~$M_J$} \\ 
Parameter & Range & Mean & Median & Std. Dev. & MAD & IQR\\
\hline
\% Unc. in $M_{p}$&0.2--195.45&13.7&9.5&15.0&4.7&10.5\\
\% Unc. in $M_{\star}$&0.76--0.76&9.6&6.6&11.64&3.53&8.2\\
\% Unc. in $P^\dagger$&0.0--0.0&3.9&0.3&10.84&0.31&1.4\\
\% Unc. in $a$&0.01--0.01&4.8&2.6&6.84&1.5&3.3\\
\enddata
\tablenotetext{\dagger}{$P$ uncertainties of zero correspond to uncertainties $< 10^{-4}$ days.}
\end{deluxetable*}

A further consideration is the effect of host star brightness on
measurement signal-to-noise and subsequently the uncertainties. Shown
in Figure~\ref{fig:corr2} are the percentage uncertainties for the
planetary masses (left column) and semi-major axes (right column) of
our exoplanet sample as a function of host star brightness. As for
Figure~\ref{fig:corr1}, the data are separated by those detected via
RV surveys only (cyan circles) and those detected via RV and transit
surveys (maroon diamonds). Although transit surveys can more
efficiently detect planets around fainter stars than RV surveys, it is
notable that the distribution of uncertainties for planet mass and
semi-major axis do not substantially differ as a function of host star
brightness. This may be due to the fact that transiting planets
provide tight constraints on the planet's period and transit epoch
thereby reducing the number of free parameters in the fit and allowing
for more precise mass and semi-major axis uncertainties. Transiting
planets also allow for more intentional RV coverage across the
planet's orbital phase which results in more accurate and precise
masses \citep{burt2018,lam2024c}.


\subsection{Effects on Dynamical Simulations}
\label{effects}

As described in Section~\ref{corr}, an impetus for improving the
planet mass and semi-major axis values lies in their critical role in
assessing the dynamical stability and evolution of planetary systems
\citep{maire2023}. Orbital dynamics has proven to be a useful tool in
predicting the possible locations of additional planets in known
systems \citep{barnes2004b,rivera2007,kopparapu2009,agnew2019}, as
well as assessing their dynamical packing
\citep{fang2013,obertas2017,kane2020b,kane2023a,obertas2023}. Dynamical
simulations have also been applied to evaluating dynamical stability
in the HZ, with application to terrestrial planet climates and the
sustainability of temperate conditions
\citep{williams2002,dressing2010,kane2012e,carrera2016,kane2019e,kane2022a,li2022e,vervoort2022}. Such
work relies upon the robustness of reported planetary system
parameters, both in terms of the measurement precision and accounting
for other possible planets and interactions
\citep{beauge2008,wittenmyer2019a}.

The precise effects of system uncertainties on dynamical analyses are
highly dependent upon the system architecture under consideration
\citep{hadden2018b,kane2023c}. By way of example, consider the
HD~141399 system, which contains four giant planets with a diversity
of star--planet separations \citep{vogt2014b}. The dynamical analysis
of the HD~141399 system provided by \citet{kane2023c} revealed the
complex mean motion resonance (MMR) structure of the system, and the
consequences for dynamical stability within the HZ. Many of the
adjacent primary MMR locations are separated by a distance of only
10\%, such as the 3:1 (0.201~AU) and 5:2 (0.227~AU) MMR locations for
planet b. As shown in Section~\ref{corr}, while the median semi-major
axis uncertainty is only 2.2\%, values can be as high as 60\%. These
semi-major axis uncertainties can have significant consequences for
the correct diagnosis of dynamical aspects, such as the apsidal
behavior of orbital evolution (libration or circulation) near MMR
locations \cite{barnes2006c,kane2014b} and can ultimately lead to
incorrect conclusions regarding the long-term stability prognosis of
planetary systems.


\section{Perturbers Within the Habitable Zone}
\label{pert}

In this section, we discuss the importance of detecting additional
gravitational perturbers in the HZ, and the RV precision required to
achieve this for HWO targets.


\subsection{Habitable Worlds Observatory Targets}
\label{hwo}

\begin{figure*}[htb!]
\begin{center}
\includegraphics[width=.9\textwidth]{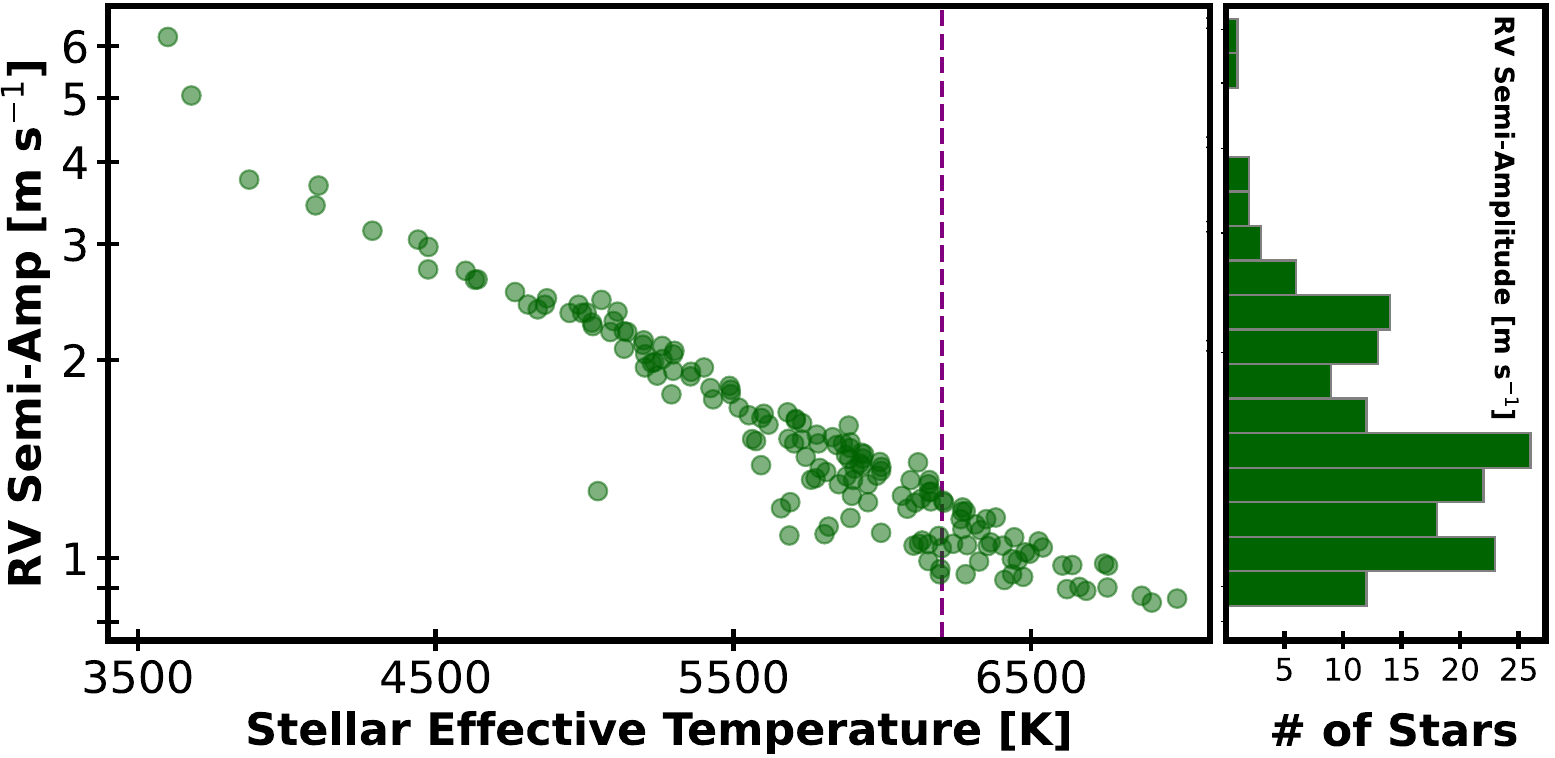}
\caption{The RV semi-amplitude of a Neptune-mass planet in the middle
  of the OHZ for each of the stars on the \citet{mamajek2024} HWO
  target list, shown as a function of stellar effective temperature
  (left panel) and as a histogram (right panel). The vertical dashed
  line indicates the Kraft break at T$_{\rm{eff}}$ = 6250~K, beyond
  which precision RV measurements become increasingly challenging due
  to the rapid rotation of the stars and the associated broadening of
  the stars' absorption lines. The majority these hypothetical HZ
  Neptunes would have RV semi-amplitudes below 1.5\ms\ making them
  challenging detections, especially at the 100+ day orbital periods
  of these HZs. Long term, dedicated EPRV survey efforts would be
  required to detect these potential perturbers and determine what
  portions of each star's HZ remains dynamically viable for Earth
  analog planets. For the hottest stars, similar surveys via precision
  astrometry could provide complementary insights into the star's HZ
  inhabitants.}\label{fig:rv}
\end{center}
\end{figure*}

The recent work by \citet{kane2024d} demonstrated the importance of
understanding the full dynamical effects of system architectures on
the possible presence of planets in the HZ. Here, we use the proposed
HWO target list provided by \citet{mamajek2024} to calculate the
signature of an additional, Neptune-mass, planet within the HZ
(described in Section~\ref{rv}) for all 164 target stars. The stars
are all relatively bright, with a V magnitude range of 0.0--7.4, and
span a broad range of spectral types, with an effective temperature
range of 3601--6990~K.

For each star, we calculated the HZ boundaries using the stellar
parameters provided by \citet{mamajek2024} and the prescription
described by \citet{kopparapu2013a,kopparapu2014}. This methodology
adopts the products from 1D Earth-based climate models to determine
the radiative balance for which surface liquid water is retained for a
planet with an Earth-based atmospheric composition. The HZ is
considered to contain two primary regions: the conservative HZ (CHZ)
and the optimistic HZ (OHZ). The inner edge of the CHZ is the boundary
where water vapor saturates the stratosphere and the oceans evaporate
entirely, and the outer edge of the CHZ is the boundary where the
CO$_2$ greenhouse effect maximizes due to increased Rayleigh
scattering \citep{kopparapu2013a,kane2016c}. The boundaries of the OHZ
are empirically determined boundaries that extend those of the CHZ,
based on evidence regarding the possible retention of surface water
early in the histories of both Venus and Mars
\citep{baker2001,kane2014e,way2016,orosei2018,kane2019d,kane2024b}. These
calculated HZ boundaries were used in our subsequent analysis for
placing additional planets around the HWO stars.


\subsection{Required Radial Velocity Observations}
\label{rv}

There are a variety of planet mass and orbits that can cause
significant impact orbital stability within the HZ, from those whose
orbits lie entirely within the HZ to giant planets on much longer
period but more eccentric orbits \citep[e.g.,][]{kane2019e}. Here, we
consider the simple case of a Neptune-mass planet ($M_p =
17$~$M_{\oplus}$) on a circular orbit in the middle of the HZ. This
particular configuration was chosen based on the recent re-analysis of
the GJ~357 system by \citet{kane2023d}, which showed that even a
10~$M_\oplus$ mass planet in the HZ can render a substantial fraction
of that region unstable for other terrestrial planets. For each of the
164 proposed HWO targets stars (see Section~\ref{hwo}), we calculated
the OHZ boundaries and the RV semi-amplitude for a Neptune-mass planet
that lies halfway between them. The results of these calculations are
shown in Figure~\ref{fig:rv}, both as a function of the host star
effective temperature (left panel) and as a histogram (right
panel). As expected, the RV semi-amplitude decreases with increasing
stellar effective temperature, since both the stellar mass and
semi-major axis of the HZ boundaries increase in that direction. There
are two M dwarfs (GJ~411 and GJ~887) for which the RV semi-amplitude
exceeds 6~\ms, but the vast majority have a RV semi-amplitude that is
less than 5~\ms.

Of all 164 stars, 59\% have an RV semi-amplitude for the injected
Neptune-mass planet that is less than the 1.5~\ms, peak of the
distribution. We note, however, that almost a quarter (24\%) of the
proposed HWO stars lie above the Kraft break \citep[\Teff $\geq$
  6250~K, $M_\star \geq 1.1$~$M_\odot$,][]{kraft1970} and are thus
likely to be fast rotators which broadens their stellar absorption
features and reduces the star's RV information content, making
precision RV measurements difficult \citep{beatty2015}. For these
stars, those located to the right of the vertical dashed line in
Figure~\ref{fig:rv}, RV observations will be unable to detect the
small RV signal imparted by a temperate, Neptune-mass planet and other
options (e.g. precision astrometry) would be required. When
considering only those stars below the Kraft break, the percentage of
targets where the injected Neptune-mass planet is less than 1.5~\ms,
decreases to 48\%. This decrease is expected as we are considering
lower temperature and therefore lower mass stars, and so a HZ
Neptune-mass planet will exert a proportionally larger RV signal than
it would on a higher mass star whose HZ is located further out. It is
further worth noting that, based on Equation~\ref{eqn:k} and for $M_p
\ll M_\star$, the RV semi-amplitude scales linearly with planet
mass. For example, a 10~$M_\oplus$ mass planet in the middle of the HZ
will result in a peak in the RV semi-amplitude distribution of
0.88~\ms\, rather than 1.5~\ms\, for the Neptune-mass case described
above.

Recovering long period, small RV semi-amplitude signatures remains
challenging in the face of stellar activity and correlated noise
\citep{narayan2005,kane2009c,wittenmyer2013a,fischer2016,langellier2021,luhn2023}. For
example, the community-wide RV challenge conducted by
\citet{dumusque2017} used data acquired with the High Accuracy Radial
velocity Planet Searcher (HARPS) spectrograph \citep{pepe2000}, into
which planetary signatures were injected. The results of that
challenge indicate that, for a star with 1.5~\ms stellar jitter, a
planetary signature with a RV amplitude of 1.5~\ms\ would require
$\sim$60 HARPS observations to achieve a robust detection, depending
on the brightness of the host star, the observing cadence adopted, and
the precision of the specific instrument used. Indeed, recent studies
of archival RV measurements by \citet{laliotis2023} and
\citet{harada2025} demonstrated that the RV sensitivity is not yet
sufficient to extract the signature of Saturn-mass planets that may be
present in the HZ for many of the proposed HWO target
stars. Additional effort will be required for a more detailed
assessment of chromospheric activity \citep{cale2021,isaacson2024a},
particularly for the coolest host stars where the HZ orbits and
overlap with stellar rotation periods
\citep{vanderburg2016b}. Although a full diagnosis of the specific
number of observations and optimum scheduling, reduction, and analysis
strategies required for all HWO stars is a complex and
multi-dimensional problem, it is clear that significant observational
resources are required to adequately uncover the signatures of
potential gravitational perturbers that yet remain in or near the HZ
of those systems. We provide specific recommendations toward achieving
the goal of detecting such planets in Section~\ref{conclusions}.


\section{Discussion}
\label{discussion}

There are several additional points of consideration to add to the
details of this paper. Since the discussion here has been largely
focused on the planetary signatures and uncertainties that arise from
RV observations, there remains the problem of an unknown orbital
inclination in the majority of these cases. The RV semi-amplitude
calculations of Section~\ref{rv} assume an edge-on inclination, which
will not be true in the vast majority of cases, resulting in more
stringent precision requirements to detect the correspondingly smaller
amplitude planetary RV signatures. Orbital dynamics for multi-planet
systems may be used to place lower limits on the inclination of the
planetary orbits, and subsequent effects on the HZ
\citep{kane2014d,korolik2023}. The coplanarity of the orbits within a
system is difficult to determine for non-transiting systems
\citep{fang2012f,ballard2016}, however, and will also influence the
calculated dynamics of the HZ \citep{kane2016d}. A further pathway
toward resolving the inclination ambiguity is the use of astrometric
observations, in particular those of the Gaia mission
\citep{perryman2014c,brown2021,feng2021} in combination with RV data
\citep{brandt2019,kiefer2021a,winn2022}. Such methodology may only be
applied for those stars that are fainter than the Gaia bright limit of
$G \approx 3$ \citep{lindegren2018}. Though Gaia is unlikely to reveal
planets within the HZ, the constraints placed on the inclination of
outer, giant planetary companions can provide valuable insight into
the inclination and dynamics of inner planets within the system. The
use of Gaia data has the additional benefit of greatly improving
stellar mass uncertainties \citep{stassun2017,fulton2017}, and thus
increasing the reliability of dynamical simulations, as discussed in
Section~\ref{uncert}. However, the inevitable outcome of such
observations will be to reveal that the true planetary masses are
indeed larger than the minimum masses inferred from the RV data alone,
further inhibiting the dynamical real estate available with the HZ of
target systems.

The calculations presented here assume a circular orbit for the
Neptune-mass planet in the HZ but
there are a broad range of orbital architectures that may influence
this region. For the known systems examined by \citet{kane2024d}, the
majority of HZ instability cases were due to giant planets in
eccentric orbits that pass in or near the HZ. However, the discussion
here is concerned with smaller, as yet undetected, planets that can
still render HZ orbits for terrestrial planets unviable. The dynamical
analysis of the GJ~357 system by \citet{kane2023d} demonstrated that
even a relatively small eccentricity of 0.1 can have a major effect on
the exclusion of other planetary orbits nearby. Thus, the results
presented here may be considered optimistic in the context of circular
orbits being the dynamically most favorable scenario.

Exoplanet characterization via RV observations will also be a crucial
component of ephemeris refinement to optimize the use of direct
imaging facilities \citep{kane2013c,kane2019b,dulz2020}. These
techniques can also be applied to determining the presence of possible
planetary and/or stellar companions to target stars at wide
separations
\citep{crepp2012b,wittrock2016,wittrock2017,kane2019b,dalba2021b,rickman2022,howell2024b}. If
such high-mass companions are present, they can play a powerful role
in the dynamical evolution of the system, and can also contribute
toward the scattering of volatiles from beyond the snow line, that
will ultimately build the volatile inventory of inner terrestrial
planets
\citep{raymond2004a,ciesla2015b,marov2018,ogihara2023,kane2024a}.

Finally, it is worth highlighting the importance of this work in the
design considerations for the next generation of direct imaging
missions. Revealing system architectures that can effectively impede
the presence of HZ planets within a system requires that replacement
targets be identified in order to avoid compromising the predicted
mission yield \citep{kopparapu2018,stark2019}. Delving into a fainter population
of stellar targets may necessitate an increased aperture to achieve
the required measurement signal-to-noise. Alternatively, the use of a
starshade as an external occulter, though technologically challenging,
can dramatically reduce the contrast ratio requirements for a
successful observation of a HZ terrestrial planet
\citep{turnbull2012,hu2021a,turnbull2021}.


\section{Conclusions}
\label{conclusions}

The preparation for high contrast, space-based direct imaging missions requires a
significant investment in the stellar characterization of potential
targets. Inadequate precursor studies carry the risk of spending
valuable mission time observing targets that are unsuitable for
achieving the stated goals of the mission. Where the goal is to
directly image terrestrial planets within the HZ, the use of
ground-based RV facilities should be sufficiently leveraged to ensure
the success of the mission. One such critical task is assessing the
viability of system HZ regions to dynamically harbor a terrestrial
planet.

Our work here has demonstrated the limitations imposed by the current
state of uncertainties for fundamental system parameters, including
the measured property of orbital period, and the derived properties of
stellar mass, planet mass, and semi-major axis. A major source of
uncertainty originates from the stellar mass uncertainties, which can
depend upon the utilized methodology
\citep{basu2012a,valcarce2013,young2014b}. For our sample of 1062
planets, the median planet mass and semi-major axis uncertainties are
11.1\% and 2.2\%, respectively, posing a challenge for subsequent
dynamical work. The main dynamical concern is the possible presence of
planets within the HZ whose \textbf{existence} serves to effectively exclude the
terrestrial planets that missions, such as LIFE and HWO, aim to
observe. 

We show that a hypothetical Neptune-mass planet in the middle of the
OHZ of the HWO stars induces an RV semi-amplitude less than 1.5~\ms
for the majority of the ExEP HWO target list. Signals of this size
remain challenging for modern RV facilities to detect, especially at
100+ day orbital periods of the HZs for most potential HWO target
stars, and and will require long-term, dedicated, observational
campaigns to uncover.  We therefore recommend two main community
actions to address the issues raised in this work.

First, we recommend that the exoplanet community invest in an update
of the stellar parameters for all known exoplanet host stars using a
self-consistent and homogeneous analysis. One potential approach is to
fit the stars' spectral energy distributions, combined with precise
distance estimates from Gaia DR3, and then using Bayesian Model
Averaging to incorporate information from multiple stellar models
\citep[see, e.g., ][]{vines2022,harada2024a}. Alternatively, high
resolution spectra for these stars can be used to derive spectroscopic
stellar parameters, either via comparisons to a grid of spectral
templates \citep[e.g.,][]{petigura2017a} or via direct spectral
synthesis that computes radiative transfer through a model stellar
atmosphere \citep[e.g.,][]{valenti1996,soto2018a}. Where possible, the
spectroscopically derived parameters should be used as constraints to
interpolate upon stellar isochrones
\citep[e.g.,][]{demarque2004,choi2016} to derive additional mass,
radius, and age ranges for the stars. As most exoplanet
characteristics are derived parameters whose value and uncertainty
rely on our knowledge of the host star's mass and/or radius, updates
to the stellar parameters will then need to be propagated through to
the planet parameters. A full re-analysis of the original time series
data that led to the detection of the planet, either the transit
photometry or RV time series, would provide the most cohesive view of
each system. However, an intermediate option would be to take the
measured quantity (either the transit depth or the RV semi-amplitude)
and re-derive the corresponding exoplanet radius or mass measurement
using the updated stellar radius or mass.

Second, we recommend the initiation of dedicated RV surveys that
target the list of the likely HWO direct imaging targets and search
for evidence of disruptive planets. Recent simulation efforts from
\citet{luhn2023} and observational efforts of the sun from
\citet{klein2024b} and \citet{ford2024} suggest that the detection of
HZ Neptunes and even super-Earths at the $\sim$10 Earth mass level is
attainable with modern RV instruments. The application of a Fischer
information metric that incorporates information about correlated
stellar noise \citep[see, e.g.,][]{cloutier2018b,gupta2024d,lam2024c}
may provide an efficient way to survey the HZs of these stars. The
design and execution of such a survey will likely require multiple
teams working in coordination, as the HWO targets span the entire sky
and cannot be comprehensively observed from a single facility. Such a
survey will also require a multi-year investment of observing
resources as the orbital periods of these HWO HZs are 100s of days
long, and establishing the stability of an RV signal over multiple
orbits is often a requirement for confirming its Keplerian nature,
particularly for those RV semi-amplitudes near the 1.5~\ms peak
described in Section~\ref{rv}. As a first step, we therefore recommend
that the exoplanet community determine the most effective and
efficient way to conduct such a survey and compare the resulting
observational needs with the current suite of accessible RV facilities
to determine what additional resources / observing support may be
required. Once such a plan has been developed and vetted by the
community, execution should begin as soon as possible to ensure that
the outer regions of the HZ are thoroughly investigated in time to
provide insights to HWO's architectural design and science
operations. The detection of these dynamical perturbers within the HZ
will not only guide the target selection of direct imaging missions
but may also guide the mission architectures if such perturbers are
found to be common place, thus requiring an increase in mission target
lists and a corresponding increase in mission aperture or duration.
The results of such a survey would also provide fundamental
demographics information that informs the occurrence of potentially
habitable planets.


\section*{Acknowledgements}

We acknowledge support from the NASA Astrophysics Decadal Survey
Precursor Science (ADSPS) program under grant No. 80NSSC23K1476.

Part of this research was carried out at the Jet Propulsion
Laboratory, California Institute of Technology, under a contract with
the National Aeronautics and Space Administration (NASA).

This research has made use of the NASA Exoplanet Archive, which is
operated by the California Institute of Technology, under contract
with the National Aeronautics and Space Administration under the
Exoplanet Exploration Program. This research has also made use of the
Habitable Zone Gallery at hzgallery.org. The results reported herein
benefited from collaborations and/or information exchange within
NASA's Nexus for Exoplanet System Science (NExSS) research
coordination network sponsored by NASA's Science Mission Directorate.

We follow the
\dataset[guidelines]{https://doi.org/10.5281/zenodo.10161527} of
NASA’s Transform to OPen Science (TOPS) mission for our open science
practices.  The software used to execute the analyses in this paper
and to generate the corresponding figures is preserved, alongside the
input data sets, on Zenodo at this
\dataset[DOI]{10.5281/zenodo.13363032}.




\end{document}